\documentclass[aps,reprint,twocolumns,pre,superscriptaddress,floatfix,notitlepage,nofootinbib]{revtex4-1}
\usepackage{amssymb,amsmath,amsfonts} 
\usepackage{mathtools}
\usepackage{physics}
\usepackage{array}
\usepackage{graphicx,epsfig,xcolor}
\usepackage[colorlinks=true,citecolor=blue,linkcolor=red,pdfauthor={Corps and Relano}]{hyperref}
\usepackage{newtxtext,newtxmath}

\newcommand{\beq}{\begin{equation}}
\newcommand{\eeq}{\end{equation}}

\begin{document}

\title{Lindbladian quantum chaos beyond classical strange attractors}

\author{\'{A}ngel L. Corps}
    \email[]{corps.angel.l@gmail.com}
\address{Institute of Particle and Nuclear Physics, Faculty of Mathematics and Physics, Charles University, V Hole\v{s}ovi\v{c}k\'{a}ch 2, 180 00 Prague, Czech Republic}
    \affiliation{Grupo Interdisciplinar de Sistemas Complejos (GISC),
Universidad Complutense de Madrid, Av. Complutense s/n, E-28040 Madrid, Spain}
    
\author{Armando Rela\~{n}o}
    \email[]{armando.relano@fis.ucm.es}
    \affiliation{Grupo Interdisciplinar de Sistemas Complejos (GISC),
Universidad Complutense de Madrid, Av. Complutense s/n, E-28040 Madrid, Spain}
    \affiliation{Departamento de Estructura de la Materia, F\'{i}sica T\'{e}rmica y Electr\'{o}nica, Universidad Complutense de Madrid, Av. Complutense s/n, E-28040 Madrid, Spain}

\date{\today} 

\begin{abstract}
We investigate the connection between classical and quantum chaos in a dissipative system of two coupled collective spins. Although the classical dynamics possesses a stable fixed-point attractor, we find pronounced signatures of chaos in the transient dynamics preceding relaxation. In the quantum model, such transient chaos is accompanied by a transition to random-matrix statistics in the relevant part of the Liouvillian spectrum and by a qualitative increase in the number of Liouvillian modes contributing to expectation values of physical observables. The transition occurs before the bifurcation at which the stable fixed point disappears and a chaotic attractor emerges. Our results show that Liouvillian quantum chaos can be associated with transient classical chaos, rather than requiring a chaotic asymptotic attractor, and provide a refinement of the Grobe–Haake–Sommers conjecture.
\end{abstract}

\maketitle

\textit{Introduction.--} The connection between classical chaos and quantum dynamics is well understood in closed quantum systems \cite{DAlessio2016,Haake2018}. When the classical dynamics become chaotic, universal Wigner-Dyson random-matrix \cite{Mehta2004} statistics emerge in the quantum spectrum \cite{Bohigas1984}, while regular or integrable dynamics are associated with Poisson statistics \cite{Berry1977}. This connection explains why and how quantum chaotic systems thermalize through the eigenstate thermalization hypothesis (ETH) \cite{Deutsch1991,Srednicki1994,Rigol2008,Polkovnikov2011,Gogolin2016,Mori2018}, and it extends to other dynamical phenomena, such as the emergence of symmetry-breaking phases of different natures \cite{Murthy2023,Corps2022PRB,Corps2023,Gomez2026}. These quantum-classical correspondences can be observed even in systems with a relatively small number of particles \cite{Emary2003,Villasenor2020,PerezFernandez2011,Mondal2020,Mondal2021,Bandyopadhyay2004}, and they provide remarkably accurate predictions of the energy values or initial conditions for which standard thermalization will ultimately fail \cite{Vidmar2016,Santos2010,Corps2026,Corps2022}.

For dissipative systems, the situation is far more complex. In the classical limit, trajectories may converge to either a fixed point, a limit cycle, or a strange attractor \cite{Ott2002,Li2022,Villasenor2024}; in the last case, trajectories display the typical features of chaos, such as positive Lyapunov  exponents \cite{Strogatz2015}. Indeed, the Grobe-Haake-Sommers (GHS) conjecture proposes a link between such classical strange attractors and random-matrix statistics in the Liouvillian superoperator that generates the time evolution from the Lindblad equation \cite{Grobe1988,Grobe1989}. However, this conjecture has been shown not to hold generally. The first counter-example of such correspondence was found in the open Dicke model \cite{Villasenor2024}. Other instances of random-matrix statistics occurring without corresponding chaotic structures in the classical dynamics have also been reported \cite{Ferrari2025,Mondal2025,Villasenor2025,Naves2026}. Although unexpected, this result is somehow logical. Random-matrix statistics is usually derived from the whole spectrum of the Liouvillian operator, whereas only a very small subset of its eigenstates ---those corresponding to eigenvalues with null real part--- give rise to the quantum version of classical attractors. Thus, the relation between classical and quantum chaos in open systems remains unsettled \cite{Sa2026}. While early studied focused on random-matrix spectral statistics \cite{Grobe1988,Akemann2019,Denisov2019,Sa2020,Li2021}, universality classes and symmetry classifications \cite{Hamazaki2020,Prasad2022,RubioGarcia2022,GarciaGarcia2022,Kawabata2023,Sa2023,Kawabata2019,Gong2018},  recent works point towards identifying quantum signatures of chaos directly from
dissipative dynamics itself \cite{Ferrari2025,Seibold2026} and its control \cite{Mondal2026arxiv}. A link between random-matrix spectral statistics and transient signatures of quantum chaos has also been proposed \cite{Mondal2026,Ferrari2025}. Finally, experimental studies of dissipative quantum chaos have recently appeared \cite{Peyruchat2025,Wold2025}.

In this Letter, we propose that random-matrix statistics in open quantum systems is linked to classical manifestations of chaos, not necessarily in the form of a strange attractor, but also in the transient dynamics leading the trajectory to a regular attractor. This is observed in a dissipative quantum system with a well-defined classical limit, and a bifurcation separating a region characterized by a stable fixed point from another in which trajectories evolve towards a strange attractor. We show that clear manifestations of classical chaos arise before reaching this bifurcation. When this happens, a small change in the initial condition implies a huge change in the time at which a trajectory gets trapped by the attractor; this phenomenon has been observed in other classical dissipative systems \cite{Lai2011}. In the quantum version of the model, random-matrix statistics emerge around the same values of the parameters at which classical manifestations of chaos in the transient dynamics appear. Furthermore, the expectation values of physical observables in the Liouvillian eigenmodes, one of the main ingredients of generalizations of the ETH to open quantum systems \cite{Hamazakiarxiv,Almeida2026,Moudgalya2019,Cipolloni2024,Roy2025,Shirai2020,Richter2025}, undergo abrupt changes around the very same time.


\textit{Model.--} We consider a quantum many-body system whose unitary dynamics is described by a Hamiltonian, $\hat{H}$, weakly coupled to an environment. The dynamics of the open quantum system \cite{Breuer2002} is encoded in its density matrix $\rho(t)$, which follows the quantum master equation $\dot{\rho}=\mathcal{L}[\rho]$, where $\mathcal{L}$ is the Liouvillian superoperator. For a Markovian environment, $\mathcal{L}$ follows the celebrated Lindblad equation \cite{Linblad1976,Gorini1976}:
\beq\label{eq:liouvillian}
\mathcal{L}[\rho]=-i[\hat{H},\rho]+\sum_{i=1}^{r}\left(\hat{L}_{i}\rho \hat{L}_{i}^{\dagger}-\frac{1}{2}\{\hat{L}_{k}^{\dagger}\hat{L}_{k},\rho\}\right).
\eeq
Here, $\hat{L}_{i}$ are the so-called jump operators describing the dissipation induced by the environment, $r$ is the number of dissipation channels, and $[\hat{A},\hat{B}]$ ($\{\hat{A},\hat{B}\}$) denotes the commutator (anti-commutator) of the operators $A$ and $B$.  The superoperator $\mathcal{L}[p]$ in Eq. \eqref{eq:liouvillian} generates the completely-positive trace preserving map $e^{\mathcal{L}t}$, and as a consequence $\rho(t) = e^{\mathcal{L}t}\rho(0)$. 

\begin{figure*}[t]
    \centering
    \includegraphics[width=0.8\textwidth]{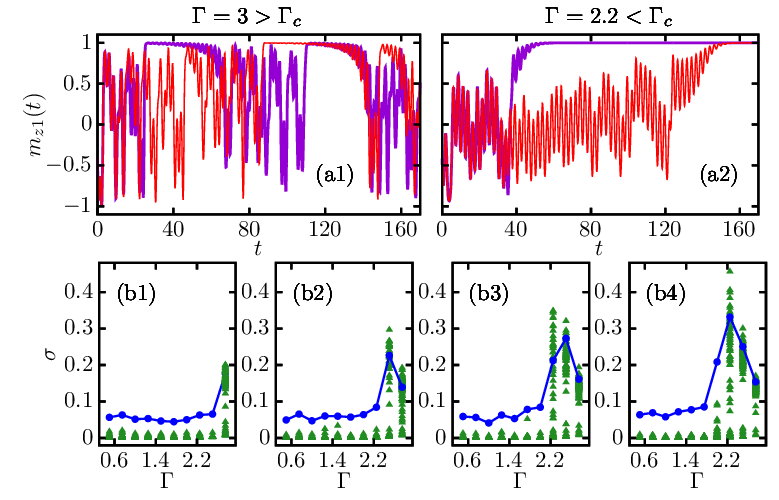}
    \caption{\textbf{Infinite-size limit: onset of classical chaos}. (a1-a2) Time evolution of the magnetization along the z-axis for the first spin subsystem. Two slightly different initial conditions are considered in both cases:  $\Gamma=2.2<\Gamma_{c}$, $\mathbf{m}(0)=(-0.04309, 0.9227, 0.9074, -0.3592, -0.4180, 0.1401)$ (purple lines) and $\mathbf{m}(0)=(-0.04286, 0.9224, 0.9068, -0.3599, -0.4193, 0.1405)$ (red lines). For (a1), $\Gamma=3>\Gamma_{c}$, and for (a2), $\Gamma=2.2<\Gamma_{c}$.  (b1-b4) Standard deviations of the normalized critical relaxation time, $t_{\mathrm{crit}}/\tau$, as a function of $\Gamma$ for initial conditions at distances $d=0.8$, $1.2$, $1.6$, and $2$ from the stable fixed point, respectively. For each $(\Gamma,d)$, 100 random initial conditions are generated, and 100 Gaussian perturbations with $\sigma=10^{-2}$ are generated around each initial condition. Solid blue lines with circles show $\sigma_d$, the standard deviation over the 100 initial conditions, while triangles show $\sigma_{\epsilon}$, the standard deviation over the perturbations of each initial condition. The increasing similarity between $\sigma_d$ and $\sigma_{\epsilon}$ signals the emergence of a fractal structure in phase space. The remaining model parameters are $\Omega=\kappa=1$. }
    \label{fig:classical}
\end{figure*}

We consider the open dynamics of two coupled spins of magnitude $S$, with Hamiltonian
\beq
\label{eq:hamiltonian}
\hat{H}=-\Omega\left(\hat{S}_{z1}+\hat{S}_{z2}\right)-\frac{\Gamma}{S}\hat{S}_{x1}\hat{S}_{x2}+\alpha(\hat{S}_{x1}+\hat{S}_{x2}).
\eeq
Here, $\hat{S}_{\beta k}$ ($\beta\in\{x,y,z\}$, $k\in\{1,2\}$) are pseudo-spin operators $\hat{S}_{\beta k}=\frac{1}{2}\sum_{\ell=1}^{N}\hat{\sigma}_{k \ell}^{\beta}$. Choosing $S=N/2$, we restrict our analysis to the sector of maximum total spin. Therefore, the thermodynamic limit, $N\to\infty$, leads to the classical limit, $S\to\infty$ \cite{Iemini2018,Postavova2026}. The coupling strength between spins $1$ and $2$ is controlled by $\Gamma$, $\Omega$ is the driving frequency of the two subsystems, and $\alpha$ is a symmetry-breaking perturbation. If $\alpha=0$, Eq. \eqref{eq:hamiltonian} is the Hamiltonian of the coupled top, which develops quantum chaos in the central part of the spectrum \cite{Feingold1983,Fan2017,Jangid2026}. We consider two dissipation channels ($r=2$), with jump operators $\hat{L}_{1}=\sqrt{\frac{\kappa}{S}}\hat{S}_{+1}$ and $\hat{L}_{2}=\sqrt{\frac{\kappa}{S}}\hat{S}_{-2}$, with $\kappa$ being the decay (or excitation) rate; with this choice, we expect dissipation to drive the system towards the central part of the spectrum. Quantum manifestations of chaos have been reported in the open version of similar models \cite{Braun1999,Postavova2026,Passarelli2025}. 

To obtain the Liouvillian spectrum, the density matrix $\rho$, acting on a Hilbert space of dimension $\mathcal{D}$, is vectorized into a state \(\lvert\!\lvert\rho\rangle\!\rangle\) in Liouville space of dimension $\mathcal{D}^2$. Denoting  $\operatorname{vec}(\rho)$ the vectorization of the density matrix by column-wise stacking its matrix elements, and using $\mathrm{vec}(\hat{A}\rho \hat{B})=(\hat{B}^{T}\otimes \hat{A})\mathrm{vec}(\rho)$, the master equation can be written as $
\frac{d}{dt}\lvert\!\lvert\rho\rangle\!\rangle
=\mathbb{L}\lvert\!\lvert\rho\rangle\!\rangle$ where
$
\mathbb{L}
=-i\left(\mathbb{I}\otimes \hat{H}-\hat{H}^{T}\otimes\mathbb{I}\right)
+\sum_i\left[
\hat{L}_i^*\otimes \hat{L}_i
-\frac{1}{2}\mathbb{I}\otimes \hat{L}_i^\dagger \hat{L}_i
-\frac{1}{2}(\hat{L}_i^\dagger \hat{L}_i)^T\otimes\mathbb{I}
\right].
$
The Liouvillian $\mathbb{L}$ is therefore a rapidly growing $\mathcal{D}^{2}\times \mathcal{D}^{2}$ matrix whose eigenvalues $\lambda_{\alpha}$ determine the decay rates and oscillation frequencies of the dynamical modes. In this formalism, $\mathbb{L}$ satisfies $\mathbb{L}\hat{R}_{\alpha}=\lambda_{\alpha}\hat{R}_{\alpha}$, where $\hat{R}_{\alpha}$ are the right eigenvectors. The Liouvillian has at least one zero-eigenvalue, the steady state (SS), corresponding to a special mode living in its kernel, $\mathcal{L}[\rho_{\textrm{SS}}]=0$, and invariant under $\mathcal{L}$, $\dot{\rho}_{\textrm{SS}}=0$. In our system of two spins of size $S$, $\mathcal{D}=(2S+1)^{2}$. We use exact diagonalization and are thus limited to $N\leq 12$.

In order to study the quantum-classical correspondence, we will consider the classical dissipative dynamics. The classical equations of motion are obtained from the equations for the expectation values of the collective spin operators using the adjoint Liouvillian. Introducing the normalized magnetizations $m_{\mu i}=\langle \hat{S}_{\mu i}\rangle/S$, and factorizing higher-order spin correlations in the classical (mean-field) limit, \(\langle \hat{S}_{\mu i}\hat{S}_{\nu j}\rangle\simeq\langle \hat{S}_{\mu i}\rangle\langle \hat{S}_{\nu j}\rangle\), one obtains a set of coupled non-linear differential equations, $\dot{\mathbf{m}}=\mathbf{F}(\mathbf{m})$, with $\mathbf{m}=(m_{x1},m_{y1},m_{z1},m_{x2},m_{y2},m_{z2})$ and $\mathbf{F}$ a vector field giving rise to
\beq\label{eq:classicaleqs}
\begin{aligned}
\dot{m}_{x1} &= \Omega m_{y1} - \kappa m_{x1}m_{z1}, \\
\dot{m}_{y1} &= -\Omega m_{x1} - (\alpha-\Gamma m_{x2})m_{z1}
-\kappa m_{y1}m_{z1}, \\
\dot{m}_{z1} &= (\alpha-\Gamma m_{x2})m_{y1}
+\kappa(m_{x1}^2+m_{y1}^2), \\
\dot{m}_{x2} &= \Omega m_{y2} + \kappa m_{x2}m_{z2}, \\
\dot{m}_{y2} &= -\Omega m_{x2} - (\alpha-\Gamma m_{x1})m_{z2}
+\kappa m_{y2}m_{z2}, \\
\dot{m}_{z2} &= (\alpha-\Gamma m_{x1})m_{y2}
-\kappa(m_{x2}^2+m_{y2}^2).
\end{aligned}
\eeq
If initially $|\mathbf{m}_{k}(0)|=1$, then $|\mathbf{m}_{k}(t)|=1$ for all $t$ as the unit Bloch sphere is an invariant surface under Eq. \eqref{eq:classicaleqs}.

\begin{figure*}[t]
    \centering
    \includegraphics[width=0.69\textwidth]{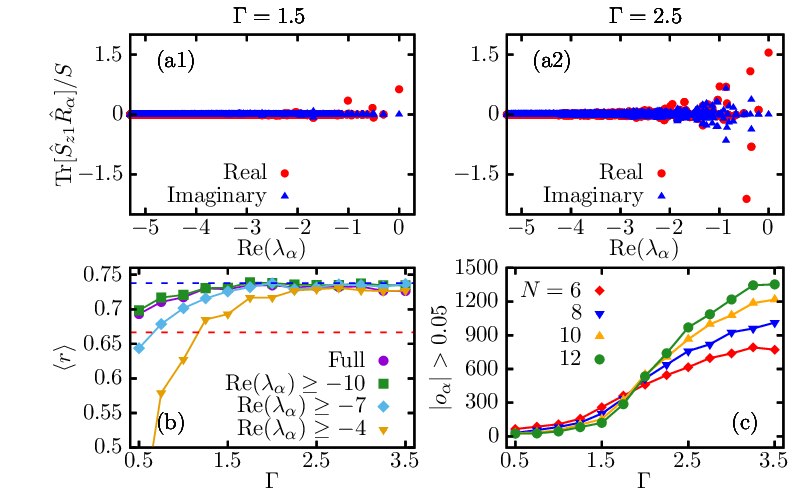}
    \caption{\textbf{Signatures of dissipative quantum chaos.} (a1-a2) Expectation values of $\hat{S}_{1z}$ in the right eigenvectors $\hat{R}_{\alpha}$ as a function of the real part of the Liouvillian eigenvalues $\lambda_{\alpha}$. Real (red circles) and imaginary (blue triangles) parts are plotted separately, for (a1) $\Gamma=1.5<\Gamma_{c}$ and (a2) $\Gamma=2.5>\Gamma_{c}$ and $N=12$. (b) Average value of the complex ratio of nearest-neighbor eigenvalues on the complex plane \cite{Sa2020} as a function of $\Gamma$ for $N=12$, for the full spectrum (purple circles) and for Liouvillian eigenvalues with different $\textrm{Re}(\lambda_{\alpha})$. The blue dashed line represents the random-matrix GinUE value $\langle r\rangle\approx 0.738$ while the red dashed line represents the Poisson average $\langle r\rangle = 2/3$. (c) Number of Liouvillian expectation values $\hat{O}=\hat{S}_{k}/S$ on the right eigenvectors $\hat{R}_{\alpha}$, $o_{\alpha}=\textrm{Tr}[\hat{O}\hat{R}_{\alpha}]$, greater than $0.05$ for different system sizes $N$. All six spin operators $\hat{S}_{k\beta}$ are accounted for in these results. The remaining model parameters are $\Omega=\kappa=1$ throughout. }
    \label{fig:quantum}
\end{figure*}

\textit{Classical analysis.--} We begin by studying the types of attractors that emerge from Eq. \eqref{eq:classicaleqs}. We find the fixed points, $\mathbf{m^*}$, such that $\mathbf{F}(\mathbf{m^*})=0$ and $|\mathbf{m^*_1}| = |\mathbf{m^*_2}|=1$, and we study their linear stability, $\delta\dot{\mathbf{m}}\simeq J(\mathbf{m}^{*})\delta\mathbf{m}$, where $J$ is the corresponding Jacobian matrix, $J_{ij}(\mathbf{m}^{*})=\eval{\partial F_{i}/\partial m_{j}}_{\mathbf{m}=\mathbf{m}^{*}}$. For $\Omega=\kappa=1$, we numerically find a bifurcation taking place at $\Gamma_{c}\approx 2.826$. For $\Gamma<\Gamma_{c}$, there always exists one (and only one) fixed point with $\textrm{Re}(j_{k})<0$ for all $k$, which is therefore stable. On the contrary, for $\Gamma>\Gamma_{c}$, all fixed points have eigenvalues with positive real parts, and therefore Eq. \eqref{eq:classicaleqs} has no stable fixed points in this region.

In Fig. \ref{fig:classical}(a1-a2) we have represented the classical dynamics of $m_{z1}=\langle \hat{S}_{z1}\rangle/S$. In each case, we select an initial condition on the unit-sphere, $\mathbf{m}(0)$, as well as a slight perturbation of this initial condition, and solve the system of differential equations Eq. \eqref{eq:classicaleqs} for each of them.

In Fig. \ref{fig:classical}(a1) we display $m_{z1}(t)$ for $\Gamma=3>\Gamma_c$ (other components of $\mathbf{m}(t)$ behave in the same qualitative way). We observe extreme sensitivity to initial conditions and no traces of periodicity. As the same qualitative behavior is observed for other values of $\Gamma>\Gamma_c$ (not shown), we conclude that this region of the parameter space is characterized by the presence of chaotic attractors. Trajectories have positive Lyapunov exponents.

In Fig. \ref{fig:classical}(a2) we show the same observable for $\Gamma=2.2 < \Gamma_c$. For this value of $\Gamma$, all trajectories evolve towards the same stable fixed point. Therefore, as the maximal Lyapunov exponent is defined $\Lambda=\lim_{t \rightarrow \infty} \lim_{\mathbf{\delta}_0 \rightarrow 0} \frac{1}{t} \ln (|\mathbf{\delta}(t))|/|\mathbf{\delta}_0|)$, all trajectories have {\em negative} asymptotic Lyapunov exponents, $\Lambda=\textrm{Re}(\Lambda_{\textrm{max}})$, where $\Lambda_{\textrm{max}}$ is the eigenvalue of the Jacobian matrix, evaluated at the stable fixed point, with the largest real part \cite{Goldhirsch1987}. Nevertheless, clear signatures of chaos are still present in this panel. Two trajectories whose initial conditions are separated by $d \approx 1.69 \cdot 10^{-3}$ in the six-dimensional phase space follow a completely different path to the same fixed point. At $\tau \sim 50$, the first one is is so close to the fixed point that its distance from it is imperceptible, whereas the second one does not reach the same stage until $\tau \sim 150$. This means that Eq. \eqref{eq:classicaleqs} with $\Gamma=2.2 < \Gamma_c$ is characterized by a chaotic {\em transient} behavior with a clear observable consequence: an enhanced uncertainty in the prediction on the relaxation time.

Fig. \ref{fig:classical} (b1)-(b4) provides a deeper insight into this feature. We display a collective portrait of $4 \cdot 10^5$ trajectories chosen in the following way: (i) We select $10$ different values of $\Gamma$; (ii) for each one, we choose four distances to the fixed point ($d=0.8$, $1.2$, $1.6$, and $2$); (iii) for each pair ($\Gamma$, $d$), we generate $100$ randomly chosen initial conditions, $\mathbf{m}_i(0)$; (iv) and for each triplet ($\Gamma$,$d$,$\mathbf{m}_i(0)$), we generate $100$ random perturbations from a Gaussian distribution with $\mu=0$ and $\sigma=10^{-2}$. We define $t_{\mathrm{crit}}$ as the time required to reach a neighborhood of radius \(10^{-4}\) around the fixed point and normalize it by the relaxation time $\tau$ predicted by linear stability theory (more details are given in Appendix \ref{sec:appendixfractal}). We quantify the heterogeneity of the resulting relaxation times through their standard deviation, $\sigma$. We use $\sigma_d$ to deonte the standard deviation of the $100$ randomly chosen initial conditions with fixed ($\Gamma$,$d$), and $\sigma_{\epsilon}$ for the standard deviation of the $100$ random perturbations with fixed ($\Gamma$,$d$,$\mathbf{m}_i(0)$). Solid lines with black dots represent $\sigma_d$. At each pair ($\Gamma$,$d$), we plot $100$ green triangles, each one displaying the value of $\sigma_{\epsilon}$ for each initial condition $\mathbf{m}_i(0)$. 
Under normal circumstances, it is expected that $\sigma_{\epsilon} \ll \sigma_d$, since $\sigma_d$ is obtained from trajectories covering a large portion of the phase space, whereas $\sigma_{\epsilon}$ comes from a very small region. 

Results in Fig. \ref{fig:classical}(b1)-(b4) indicate that the structure of the phase space is very complex. Near the fixed point (Fig. \ref{fig:classical}(b1), corresponding to $d=0.8$), $\sigma_d$ is small and $\sigma_{\epsilon}$ is clearly smaller, except for $\Gamma=2.75$. In this case, both magnitudes are larger and $\sigma_d \approx \sigma_{\epsilon}$ for a significant number of initial conditions. This is a signature of a fractal geometry in the phase space \cite{Ott2002}: the heterogeneity of a hypersphere with radius $d=0.8$ centered at the fixed point is very similar to the heterogeneity of a very small hypersphere centered at a particular initial condition. Fig. \ref{fig:classical}(b2)-(b4) shows that the larger the distance $d$ to the fixed point, the lower the value of $\Gamma$ at which the signatures of fractal behavior emerge. From Fig. \ref{fig:classical}(b4), we conclude that huge sensitivity to initial conditions and fractality in classical phase space start to occur at $\Gamma \gtrsim 2$ if $d=2$. This is compatible with the chaotic behavior displayed in Fig. \ref{fig:classical}(a2), with $\Gamma=2.2$. Furthermore, for $\Gamma=2.75$, signatures of a fractal structure are visible through almost the entire phase space.

From these results, we draw the following picture. For $\Gamma \gtrsim 2$, clear signatures of classical chaos emerge far from the stable fixed point. Upon increasing $\Gamma < \Gamma_c$, these signatures expand throughout the whole phase space. We infer that the whole phase space becomes complex at $\Gamma=\Gamma_c$, the basin of attraction of the fixed point disappears and strange attractors emerge. This implies that the onset of classical chaos occurs for dissipation strengths $\Gamma$ well below the critical bifurcation value $\Gamma_{c}$. This is despite the fact that for these lower values of $\Gamma$ the fixed point is stable, suggesting that in the thermodynamic limit, a chaotic attractor is not a necessary condition for chaos.

\textit{Quantum analysis.--} We now move on to the analysis of the quantum signatures of chaos. For a given initial state $\rho(0)$, the time evolution of an observable, $\langle \hat{O}(t)\rangle=\textrm{Tr}[\hat{O}e^{\mathcal{L}t}\rho(0)]$ follows the spectral decomposition
\beq\label{eq:Ot}
\langle \hat{O}(t)\rangle=\sum_{\alpha}e^{\lambda_{\alpha}t}\textrm{Tr}[\hat{O}\hat{R}_{\alpha}]\textrm{Tr}[\hat{L}_{\alpha}^{\dagger}\rho(0)],
\eeq
where $\hat{L}_{\alpha}$ are the Liouvillian left eigenvectors, $\mathcal{L}^{\dagger}\hat{L}_{\alpha}=\lambda_{\alpha}^{*}\hat{L}_{\alpha}$ (here, orthonormality of $\hat{L}_{\alpha}$ and $\hat{R}_{\alpha}$ is assumed).

In the thermodynamic limit (TL) of our model, this expression must reproduce the large heterogeneity in relaxation times shown in Fig. \ref{fig:classical}. In its quantum version, the relaxation time of a particular initial condition is given by the Liouvillian eigenvalue with the largest real part, $\lambda_{\alpha}^{\textrm{max}}$, among those with a relevant {\em population} $c_{\alpha}^{\textrm{max}} = \textrm{Tr} [\hat{O}\hat{R}^{\textrm{max}}_{\alpha}]\textrm{Tr}[\hat{L}^{\dagger, \textrm{max}}_{\alpha} \rho(0)]$, since eigenvalues with (almost) null populations play no relevant role. This suggests that the only way different initial conditions with similar values of $\langle \hat{O}(0)\rangle$ can have very different relaxation times is through $\lambda_{\alpha}^{\textrm{max}}$ changing abruptly upon introducing a slight perturbation in the initial value $\langle \hat{O}(0)\rangle$. If there is a large set of different eigenstates with large values of $|o_{\alpha}|\equiv |\textrm{Tr}[\hat{O}\hat{R}_{\alpha}]|$, distributed over a wide range of $\textrm{Re}(\lambda_{\alpha})$, it seems easy to generate initial conditions with similar values of $\langle \hat{O}(0)\rangle$ and different values of $\lambda_{\alpha}^{\textrm{max}}$. However, if very few eigenvectors have significant values of $|o_{\alpha}|$, this task seems much more difficult.

In Fig. \ref{fig:quantum} (a1)-(a2), we represent $o_{\alpha}$ for  $\hat{O}=\hat{S}_{z1}/S$ and two values of $\Gamma<\Gamma_{c}$. For $\Gamma=1.5$, $o_{\alpha}$ is only significantly non-zero for very few eigenmodes. On the other hand, for $\Gamma=2.5$, the number of eigenmodes satisfying this condition is much larger. An interesting feature of both cases is that $|o_{\alpha}|$ is very small for $\textrm{Re}(\lambda_{\alpha}) < -4$ (the same occurs for other observables). This suggests that only a small part of the Liouvillian spectrum, close to the steady state, bears observable dynamical consequences.

To obtain a proper measure of the complexity of such $o_{\alpha}$, in Fig. \ref{fig:quantum}(c) we have represented the number of eigenmodes such that $|o_{\alpha}|>0.05$ (the bound is arbitrary; similar results are obtained for other bounds). This marks two significantly different behaviors: for $\Gamma\lesssim 1.75$, the number of modes fulfilling this condition {\em decreases} with the number of particles $N$. On the contrary, for $\Gamma \gtrsim 2$, the number of contributing modes increases with $N$. This result suggests a qualitative change in the properties of the coefficients $o_{\alpha}$ around the same value of $\Gamma$ at which transient chaos appears in the classical version of the model. Only for $\Gamma \gtrsim 2$ do the results become compatible with a sufficiently large number of contributing eigenmodes in the thermodynamic limit, coinciding with the onset of chaotic manifestations in the classical dynamics.

Finally, we study whether this qualitative change is translated into spectral statics.
We resort to the distribution of the ratio of nearest-neighbor level spacings \cite{Atas2013}, recently generalized to dissipative systems \cite{Sa2020}. We avoid the more traditional level spacing distribution to prevent the complications inherent in the unfolding procedure \cite{Gomez2002,Corps2021}. Given the Liouvillian eigenvalues $\lambda_{\alpha}$, we consider the ratio 
\beq
z_{\alpha}=r_{\alpha}e^{i\alpha}=\frac{\lambda_{\alpha}^{\textrm{NN}}-\lambda_{\alpha}}{\lambda_{\alpha}^{\textrm{NNN}}-\lambda_{\alpha}},
\eeq
where NN indicates the nearest neighbor and NNN is the next-to-nearest neighbor. For the 2D Poisson distribution, $\langle r\rangle_{\textrm{2DP}}=2/3$, while for the Ginibre unitary ensemble (GinUE), $\langle r\rangle_{\textrm{GinUE}}\approx 0.74$. We have plotted the average absolute value ratio, $\langle r\rangle$, as a function of $\Gamma$ for $N=12$ in Fig. \ref{fig:quantum}(b), for different spectral regions. If we restrict ourselves to the region with significant values of $|o_{\alpha}|$, we observe that our system becomes fully chaotic around $\Gamma \approx 2.25$. This is before the classical bifurcation point, but around the same value of $\Gamma$ at which clear signatures of classical transient chaos appear and within the region in which the number of modes with significant values of $|o_{\alpha}|$ increases with the number of particles. If we also consider fast-decaying modes with negligible values of $|o_{\alpha}|$ in the spectral statistics, the transition to chaos happens clearly before, at $\Gamma \gtrsim 1.5$.

\textit{Discussion.--} Our results allow us to draw the following conclusions. In our model, there is a strong link between the classical manifestations of chaos in physical observables and the expectation values of the same observables in the right eigenvectors of the Liouvillian operator. The key elements of this link are a large number of eigenvectors with significantly high observable expectation values and an increase in this number with system size. This feature is present when the classical manifestations of chaos consist in a strange attractor and also when there is a regular attractor instead, but the transient dynamics is chaotic. Regarding spectral statistics, which is considered a paradigmatic signature of chaos, the same link holds true if statistics is computed only from the spectral region with significant values of expectation values of physical observables in the Liouvillian eigenmodes. Thus,
we propose a reformulation of the GHS conjecture originally conceived in the 80s: {\em random-matrix spectral statistics emerge in the spectral region of the Liouvillian superoperator close to the steady state when the classical analog displays observable consequences of chaos, either in the transient or in the long-term dynamics}. This reformulation of the GHS conjecture is in agreement with the interpretation in Ref. \cite{Mondal2026} in terms of the role played by transient dynamics in dissipative chaos.

In our model, the transition to chaotic spectral statistics occurs at considerably lower coupling strengths, $\Gamma$, if wider spectral regions are considered. 
The chaotic nature of very fast-decaying Liouvillian eigenmodes might also have a classical analog, but with much weaker consequences. For example, at $\Gamma \gtrsim 1$ we have found some trajectories with small perturbations moving away exponentially, but for a period too short to draw a meaningful conclusion and with no measurable consequences in the relaxation times. Another important remark is that long-range spectral statistics is required in closed quantum systems to correctly identify full chaos: there are many examples of closed quantum systems showing Wigner-Dyson spectral statistics and deviations from full chaos, usually diagnosed through the Thouless time \cite{Suntajs2020,Sierant2020}, energy \cite{Serbyn2017}, or frequency \cite{Corps2021,Corps2020PRB}. This suggests that spectral statistics built from nearest neighbor and next-to-neighbor level spacings alone might overestimate the degree of chaos. Thus, more sophisticated tools, such as the dissipative spectral form factor and related measures \cite{Li2021,GarciaGarcia2023}, may be also needed in open quantum systems to establish a strong link between classical and quantum dynamics.

\begin{acknowledgments}
A. L. C. and A. R. acknowledge financial support from the Spanish grant PID2022-136285NB-C31 funded by Ministerio de Ciencia e Innovacion / Agencia Estatal de Investigación
MCIN/AEI/10.13039/501100011033 and FEDER “A Way of
Making Europe”. A. L. C. also acknowledges support from the Czech Science Foundation under project No. 25-16056S and the JUNIOR UK Fund project carried out at the Faculty of Mathematics and Physics, Charles University. 
\end{acknowledgments}

\appendix
\section{Algorithm for geometrical complexity of classical phase space}\label{sec:appendixfractal}
    
To characterize the structure of the phase space surrounding the stable fixed point, we analyze the distribution of relaxation times for ensembles of initial conditions at a fixed distance from the stationary state. For each value of \(\Gamma\), we randomly generate 100 initial conditions at a prescribed distance \(d\) from the stable fixed point, allowing a tolerance of \(1/200\), i.e., \(d(0)\in[d-1/200,d+1/200]\). We consider four distances, \(d=0.8,1.2,1.6,\) and \(2\). For each trajectory, we determine the critical time \(t_{\mathrm{crit}}\) required for the trajectory to enter a hyperspherical neighborhood of radius \(10^{-4}\) around the stable fixed point. To compare the nonlinear relaxation dynamics across different parameters and initial distances, we normalize this time by the relaxation time predicted by linear stability theory. Specifically, linearizing the equations of motion around the stable fixed point gives a stability matrix whose eigenvalue with the largest real part, \(\lambda_{\mathrm{max}}\), determines the slowest asymptotic relaxation mode. Within the linear approximation, the distance from the fixed point evolves as \(d(t)=d(0)e^{\mathrm{Re}(\lambda_{\mathrm{max}})t}\), yielding the characteristic relaxation time

$$
\tau=\frac{\ln[10^{-4}/d(0)]}{\mathrm{Re}(\lambda_{\mathrm{max}})}.
$$

Since the fixed point is stable, \(\mathrm{Re}(\lambda_{\mathrm{max}})<0\), and hence \(\tau>0\). We therefore consider the dimensionless relaxation time \(t_{\mathrm{crit}}/\tau\), which equals unity when the actual nonlinear relaxation time coincides with the prediction of the linearized dynamics. The standard deviation \(\sigma\) of \(t_{\mathrm{crit}}/\tau\) over the ensemble of 100 initial conditions, that we denote \( \sigma_d \), provides a measure of the heterogeneity of the phase-space region at distance \(d\) from the fixed point: small \(\sigma_d\) indicates that trajectories originating in this region exhibit similar relaxation times, whereas large \(\sigma_d\) indicates strong sensitivity of the relaxation time to the precise location of the initial condition.

We further probe the fine-scale structure of this heterogeneity by perturbing each of the 100 reference initial conditions independently. For every reference trajectory, we generate 100 nearby initial conditions by adding random Gaussian perturbations with zero mean and standard deviation \(0.01\). For each perturbed trajectory, we again calculate the normalized relaxation time \(t_{\mathrm{crit}}/\tau\) and its standard deviation. The resulting \(\sigma\), that we denote \( \sigma_{\epsilon}\), therefore characterizes the sensitivity of the relaxation time to perturbations on a much smaller phase-space scale than that associated with the full ensemble at fixed distance \(d\). Comparing the standard deviations obtained from the two ensembles provides a measure of the scale dependence of the phase-space heterogeneity. In particular, if the fluctuations observed within small neighborhoods of individual initial conditions are comparable to those obtained across the entire region at fixed distance \(d\), the heterogeneity persists upon probing progressively smaller scales. Such scale-independent structure is consistent with a fractal-like organization of the relevant phase-space region, although a direct determination of a fractal dimension would be required to establish fractality in the strict mathematical sense.

\end{document}